\documentclass{ws-p8-50x6-00}
\usepackage{amsmath}
\usepackage{epsfig}

\def\Journal#1#2#3#4{#1{\bf #2} (#4) #3}
\def\NPB{{\em Nucl.\ Phys.} B}

\def\PLB{{\em Phys.\ Lett.}  B}

\def\PRD{{\em Phys.\ Rev.} D}

\def\JHEP{{\em JHEP }}

\def\NPPS{{\em  Nucl.\ Phys.\ Proc.\ Suppl.\ }}

\def\out{\mbox{\scriptsize out}}
\def\Ko{K_{\out}}
\def\cM{{\cal {M}}}
\begin{document}

\title {Hard QCD: some results}

\author{G.Marchesini}

\address{Universit\`a degli Studi di Milano-Bicocca and 
INFN, Sezione di Milano, Italy}

\maketitle

\abstracts{I recall some points on the present status of QCD results in
  the short distance regions and illustrate the case of event shape in
  QCD radiation}

\section{Hard QCD studies}
QCD studies have produced successful results, specially for
observables involving short distance regions, for over a quarter of
century.  High accuracy calculations of hard distributions are
possible and often one is able to make quantitative predictions which
are strongly constrained.  Calculations are often difficult but there
are constant improvements in techniques and this is always associated
to a better understanding of QCD features. A consequence of these
successful results is that QCD is often considered an established and
understood theory.  Actually, there are a variety of unsolved
theoretical and experimental questions in QCD.  They range from the two
extreme regions.
On one side we still need to understand how quark and gluon elementary
fields change into hadrons at large distances. Since we think we
almost dominate the short distance regions by perturbative (PT)
methods, there are attempts to extend these methods and even the
language to the study of non perturbative (NP) corrections. This is an
approach to large distances from the ``short distance window''.
Going in the other extreme region, the standard understanding is that
in the ultraviolet region the theory becomes free and no structures
are present. But, again, this is based on PT studies. Actually, it has
been argued that short distance structures could be present in QCD
\cite{ZV}.

Hard QCD is systematically and analytically studied by
resummation of the PT expansion.  This procedure starts from free
quarks and gluons and reconstructs their interaction by using the
point-like vertices in the QCD Lagrangian.  It is reassuring to
discover afterwards that quarks and gluons cannot be free at large
distances, although this region it is not accessible to PT methods. In
order to resumm PT expansions for physical observables one needs in
general to know multi-parton distributions. Of course we need
approximations. I recall that the present PT status of the art for
multi-parton distributions concerns: collinear singularities, soft
singularities, fixed order exact matrix elements.

{\bf 1. Collinear singularities.} In general, all leading and next to
leading contributions to the multi-parton distributions are obtained
by generalizing of the DGLAP evolution equation \cite{dglap} (for a
generalization to ``exclusive'' multi-parton distributions see
\cite{gen-func}). They are given in terms of the anomalous dimensions
which are known to leading and next-to-leading order.  Recently there
are attempts to compute NNNL contributions \cite{3gamma}. Singular
contributions to multi-parton distributions require a collinear
cutoff.  Universality of collinear singularities allows a unified
description of hard distributions for all processes such as
lepton-lepton, lepton-hadron and hadron-hadron collisions.

{\bf 2. Soft singularities.} Also in this case all leading (and some
next-to-leading) contributions to the multi-parton distributions are
known.  They are given in terms of factorized soft radiation from hard
emitters \cite{gen-func}. In the case of two or three emitters it is
possible to factorize the soft radiation at two-loop \cite{2soft}. For
more emitters the two soft parton factorization requires the planar
approximation (leading order in $1/N_c$). I recall the most important
results and features of QCD we learned in the 80th and 90th:
\begin{itemize}
\item coherence of soft radiation;
\item colour connection and angular ordering;
\item gluon multiplication and hadron multiplicity;
\item hump-backed plateau;
\item jet-finding algorithms in $e^+e^-$, DIS and hadron collisions;
\item LHPD and colour preconfinement;
\item heavy quark distribution and radiation;
%\item EEC correlation;
\item jet-shape distribution.
\end{itemize}
All these results, together with the known leading (and
next-to-leading) contributions allows one to formulate multi-parton
distributions in terms of a coherent-branching algorithms.  This is
the bases of most of the present Monte Carlo simulations for hard QCD
(and beyond QCD) processes, see \cite{MC}.
  
An other area in which soft physics is involved is the small-$x$
region in DIS and hadron collisions \cite{smallx}. Here soft
``exchanged'' gluon are involved rather than the soft emitted ones, as
in the previous item. Enormous progresses have been made recently.
These results are however less constrained by PT analysis since NP
effects enter at all hard scales \cite{smallxNP}. Further progresses in
this area definitively requires further developments in our QCD
understanding beyond PT expansions.
 
{\bf 3. Exact matrix element results.} Large technical progresses have
been made both in analytical calculations \cite{high-acc-calc} and in
numerical programs \cite{high-acc-num}. These results are essential in
order to make quantitative predictions away from the soft or collinear
regions of phase space.  This is achieved by matching these exact
fixed order results with the ones from resummations described in the
previous two items \cite{PTstandards}. Similar matching procedure are
now developed on a systematically bases in Monte Carlo simulations
\cite{matchingMC}.

In the following I will illustrate the QCD calculations of jet-shape
observables which are collinear and infrared safe (CIS).  They are
expressed as linear combinations of the momenta of emitted hadrons so
that their values do not change if two momenta are replaced by their
sum when the two are collinear or when one is soft.  For these CIS
distributions one reaches the highest accuracy available in PT
calculations. This is due to the fact that the cutoff of collinear and
infrared singularities cancels and all PT terms are finite. Near the
phase space boundary the cancellation are partial and logarithmically
enhanced contributions enter the PT terms which needs to be resummed.
The particular interest of these distributions is not only that they
can be computed at the highest PT accuracy, but that they have large
$1/Q$-power corrections which are of NP origin. These corrections are
phenomenologically important. Once we have a reliable PT estimate,
with the help of data we can then constrain the size and hopefully the
nature of these NP corrections.

After a brief review of the results for event shapes in $e^+e^-$
annihilation, I discuss how these results can be extended to jet shape
distributions in hadron-hadron and lepton-hadron collisions. This will
open the way to the study of further multi-jet observables.

\section{Hadron emission in $e^+e^-$ annihilation}
Two-jet event shapes (such as thrust $T$, broadening $B$, heavy-jet
mass $M_H$ and $C$-parameter) have been intensively studied both
theoretically and experimentally.  The state-of-the-art level of PT
analysis of two-jet observables consists of \cite{PTstandards}:
resummation of all double- (DL) and single-logarithmic (SL) enhanced
contributions (due to multiple radiation of soft and collinear
partons); matching of the resummed result with the exact fixed order
result.  To reach such an accuracy one needs: soft matrix elements at
two-loop order \cite{2soft} and reconstruction of the running coupling
at the proper scale; factorization and exponentiation of observable
and kinematical constraints; proper treatment of hard intra-jet and
soft inter-jet radiation.

Besides the pure PT contributions, experimental data (see for
instance~\cite{JADE}) have revealed the presence of power suppressed
$1/Q$ corrections ($Q$ is the $e^+e^-$ center-of-mass energy), which
emerge as a shift both to mean values and differential distributions.
It is widely believed~\cite{NPgeneral} that such contributions are of
NP origin and arise from the running of the coupling into the infrared
domain.

A systematic way to deal with power corrections is provided by the
dispersive approach \cite{DMW}, in which a prescription is
given to extend the QCD coupling into the confinement region. 
In terms of this dispersive coupling, the typical large distance
contribution to an event shape $V$ can be factorized in an observable
dependent coefficient $c_V$ and a universal NP parameter
$\alpha_0$. More precisely, in any `linear' shape variable $V$ the
dependence on the rapidity $\eta_i$ and transverse momentum $k_{ti}$
of emitted hadron $i$ factorizes,
\begin{equation}
V=\sum_i k_{ti}\> f_V(\eta_i)\>,
\end{equation} 
so that the NP contribution to $V$ becomes
\begin{eqnarray}
\label{eq:deltaV}
&\delta V=C_F\>c_V\>\lambda^{\mbox{NP}},\>\qquad 
c_V=\int d\eta \>f_V(\eta)\>,\nonumber\\ 
&\lambda^{\mbox{NP}}=\frac{\mu_I}{Q}\frac{4\cM}{\pi^2}
\left(\alpha_0(\mu_I)+O(\alpha_s(Q))\right)\>,
\qquad \alpha_0(\mu_I)=\frac{1}{\mu_I}\int_0^{\mu_I}dk\>\alpha_s(k)\>.
\end{eqnarray}
Here $\cM$ is a known coefficient, the Milan factor \cite{milan},
which takes into account the non complete inclusiveness of the
observable. The NP parameter $\alpha_0$ is the average of the running
coupling in the region $k\leq \mu_I$ and measures the interaction
strength in the confinement region. The $O(\alpha_s(Q))$ piece in
\ref{eq:deltaV} is needed in order to merge PT and NP results in a
renormalon free manner and ensures that the final answer is
independent of the infra-red matching scale $\mu_I$.  The NP parameter
$\alpha_0(2\mbox{GeV})$ has been measured and appears to be consistent
with the universality hypothesis within a 10-20\% accuracy, see \cite{SZ}.
%, as shown
%in Fig.~\ref{fig:alfa0}.

%\begin{figure}[ht]
%\epsfig{file=cont-09-1999-bw.eps
%,width=0.5\textwidth}   
%\caption{$2$-$\sigma$ contours for fits to various two-jet 
%observables~\cite{SZ}.  Solid curves indicate fits to distributions,
%while dashed lines indicate fits to mean values.
%\label{fig:alfa0}}
%\end{figure}

Only very recently these techniques have been extended to the case of
three-jet observables in the near-to-planar region.  In particular the
mean values and distributions of the thrust minor $T_m$ \cite{Tm} and
the $D$-parameter \cite{Dpar} have been computed. Both of them are a
measure of QCD radiation out of the event plane.  This study has
revealed an unexpectedly rich geometry and colour dependence both of
the PT result and of the NP corrections.  The crucial point is the
rapidity dependence of the observables.  Actually, $T_m$ is
independent of rapidity, while $D$ is exponentially damped.  On the PT
side, this implies that hard parton recoil affects the observable in
the $T_m$ case, while it does not contribute to $D$.  Therefore the
$T_m$ distribution is sensitive not only to the underlying hard event
geometry (the angles between the jets), but also on its colour
configuration through the kinematical constraints which define the
event plane.

As far as NP corrections are concerned, one finds that the shift to
the $D$ distribution is simply geometry dependent, while the $T_m$
distribution is also squeezed, since the shift depends logarithmically
on $T_m$. This behaviour (already encountered for the $B$ in two-jet
events) results from a complicated interplay between PT and NP
effects.

The same physical features appearing in the NP shift to the $T_m$
distribution are also present in hadron-hadron and
lepton-hadron collisions, as we are going to discuss in the following.

\section{Out-of-plane radiation in hadron collisions}
We consider hadron collisions and study an observable similar to
$T_m$~\cite{Kouthh}.  We select events in which a $Z_0$ is produced
with large transverse momentum. We can then define an `event plane' as
the one containing the $Z_0$ momentum and the beam axis and introduce
a measure of out-of-plane radiation
\begin{equation}
\label{eq:Kohh-def}
\Ko={\sum_h} |p_h^{\out}|\>.
\end{equation}
Here $p_h^{\out}$ is the out-of-plane momentum of the hadron $h$ and
the sum extends to all hadrons with rapidity in the range
$|\eta_h|<\eta_0$, in order to avoid measurements in the beam region.
At Born level one has two incoming partons $p_1$ and $p_2$ and an
additional hard parton $p_3$ recoiling against the vector boson $q$.
%(see Fig.~\ref{fig:hadron}a).
%
%\begin{figure}[ht]
%    \epsfig{file=hadron.eps,width=0.8\textwidth}
%   \caption{
%A particular Born configuration 
%%
%(a) for the process $p\bar p\to Z_0+jet$
%%
%(b) for lepton-hadron scattering.
%\label{fig:hadron}}
%\end{figure}
%
The event-plane definition gives rise to the conservation law
\begin{equation}
\label{eq:plane-hh}
p_3^{\out}+\sum_i k_i^{\out}=0\>,
\end{equation}
so that only $p_3$ can take an out-of-plane recoil, while the remaining
hard momenta are fixed in the event plane.

The main difference between hadronic and $e^+e^-$ collisions is the
presence of initial state radiation. Due to coherence of QCD
radiation, its contribution can be factorized giving rise to the
standard parton density function for the incoming partons $p_2,p_3$.
However, while in the total cross section the hard scale is the $Z_0$
hardness, here the scale is $\Ko$. The PT distribution is essentially
given by the product of the initial state parton distributions
$P_{in}(\Ko)$ and a CIS ``radiation factor'' 
\begin{equation}
\label{eq:Kohh-dist}
\Sigma(\Ko)\sim P_{in}(\Ko)\cdot e^{-R(\Ko)}\cdot S(R')\>.
\end{equation}
The radiator $R$ is the same as that of $T_m$ in $e^+e^-$, it resums
all DL contributions and accounts for SL effects due to hard intra-jet
and soft inter-jet radiation.  It is given in terms of three hard
parton `antennae', each one proportional to the colour charge $C_a$ of
emitting parton $p_a$ ($C_a$ equals $C_F$ for a quark and $C_A$ for a
gluon)
\begin{equation}
\label{eq:Radiator}
R(\Ko)=\sum_{a=1}^3 C_a\int_{\Ko}^{Q_a}
\frac{dk}{k}\frac{2\alpha_s(2k)}{\pi}
\ln\frac{Q_a}{2k}
\simeq \sum_a C_a\frac{\alpha_s}{\pi}\ln^2\frac{Q_a}{2\Ko}\>.
\end{equation}
The hard scales $Q_a$ are determined essentially by soft radiation at
large angles, which is a characteristic of multi-jet observables. In
particular, apart from a factor due to hard collinear splitting, the
scale for the quarks is the invariant mass of the $q\bar q$ system,
for the gluon is its invariant transverse momentum with respect to the
$q\bar q$ pair.  The SL function $S$ depends on the logarithmic
derivative of the radiator $R'$ and accounts for all effects coming
from multiple secondary radiation.

The NP correction is proportional to the same parameter
$\lambda^{\mbox{NP}}$ which enters the power corrections to $e^+e^-$
event shapes in \ref{eq:deltaV} and is given by
\begin{equation}
\label{eq:Kohh-shift}
\delta \Ko=\frac{2}{\pi} \lambda^{\mbox{NP}}
\left(C_1(\eta_0-\eta_3)+C_2(\eta_0+\eta_3)+
C_3\ln\frac{Q_t}{|p_3^{\out}|}\right)\>,
\end{equation}
where $Q_t$ is the transverse momentum of the emitted boson.  Such a
shift arises from integration over the rapidity of emitted gluons
(see eq.~\ref{eq:deltaV}). Since $\Ko$ is independent of
rapidity, one has to carefully consider the effective rapidity cutoff.
For radiation from the incoming partons $p_1$ and $p_2$ it is given by
the distance between $\eta_3$ (the rapidity of $p_3$) and the
experimental resolution $\eta_0$. On the contrary, for a NP gluon
emitted from $p_3$, it is the hard parton recoil momentum $p_3^{\out}$
which provides the needed cutoff.  Moreover, since $p_3$ always takes
recoil, one has $p_3^{\out}\sim \Ko$. This interplay between PT and
NP emissions makes the shift logarithmically dependent on $\Ko$.

Unfortunately, this is not the end of the story. In hadronic
collisions one has to add a soft contribution due to the beam remnant
interactions. This is the so-called `soft underlying event', which was
systematically studied and introduced for the first time in the
analysis of the `pedestal height' in hadronic jet
production~\cite{pedestal}.  Therefore the analysis of $\Ko$ is also
important to understand the physics of soft collisions.

\section{Out-of-plane radiation DIS}
Even in DIS one can define an event plane and measure the out-of-plane
radiation. In the Breit frame, we define the thrust major $T_M$ in
analogy with $e^+e^-$ annihilation
\begin{equation}
\label{eq:TM-DIS}
T_M Q=\max_{\vec n_M\cdot\vec n=0}{\sum_h} |\vec{p}_h\cdot \vec n_M|\>,
\end{equation}
where $\vec n$ and $Q^2=-q^2$ are the Breit axis and the virtuality of
the exchanged boson $q$ respectively. As in the previous case, the sum
extends to all hadrons not in the beam direction (with $\eta<\eta_0$).
At Born level, we have one incoming parton $p_1$ which is struck by a
vector boson $q$ and produces two hard large angle partons $p_2$ and
$p_3$ recoiling one against each other. % (see Fig.~\ref{fig:hadron}b).
We define $\Ko$ again as the cumulative out-of-event-plane momentum,
where the event plane is the one containing $\vec n$ and the $T_M$
axis.  The event-plane definition implies the following kinematical
constraint
\begin{equation}
\label{eq:plane-DIS}
p_2^{\out}+\sum_{i\in U} k_i^{\out}=p_3^{\out}+\sum_{i\in D} k_i^{\out}\>,
\end{equation}
with U (D) the region containing parton $p_2$ ($p_3$).

The PT $\Ko$ distribution has the same form as
\ref{eq:Kohh-dist}, with $P_{inc}(\Ko)$ the structure function of the
incoming parton, again at the scale $\Ko$. The radiator turns out
to be the same as in $e^+e^-$ and in hadron-hadron collisions, due to
universality of soft and collinear radiation.  However the exact
functional form of the SL function $S$ differs between these, due to
different event-plane kinematics.

The NP shift is similar to the one in \ref{eq:Kohh-shift}
\begin{equation}
\label{eq:KoDIS-shift}
\delta \Ko=\frac{2}{\pi}\lambda^{\mbox{NP}}\left(
 C_1\ln\frac{Q_1^{\mbox{NP}} e^{\eta_0}}{Q}+
C_2\ln\frac{Q_2^{\mbox{NP}}}{|p_2^{\out}|}
+C_3\ln\frac{Q_3^{\mbox{NP}}}{|p_3^{\out}|}\right)\>,
\end{equation}
with $Q_a^{\mbox{NP}}$ the proper NP hard scales, which are
proportional to the ones entering the PT radiator.  The crucial
difference is that here both $p_2$ and $p_3$ are not fixed in the
event plane. This implies that the way the NP correction in
\ref{eq:KoDIS-shift} affects the PT distribution is very different
according to which phase space region is considered. Namely one has
the two regimes
\begin{itemize}
\item{$\alpha_s\ln^2\Ko/Q\gg 1$:} one has well developed QCD radiation,
 so that all $p_a\>(a=2,3)$ take recoil and
$p_a^{\out}\sim \Ko$. This gives rise to a logarithmically enhanced shift
\begin{equation}
\label{eq:caso1}
\delta \Ko\sim 
 C_1\ln\frac{Q_1^{\mbox{NP}} e^{\eta_0}}{Q}+
C_2\ln\frac{Q_2^{\mbox{NP}}}{\Ko}
+C_3\ln\frac{Q_3^{\mbox{NP}}}{\Ko}\>.
\end{equation}
\item{$\alpha_s\ln^2\Ko/Q\ll 1$:} 
radiation from one hard parton dominates. According to the event-plane
kinematics in \ref{eq:plane-DIS}, when PT radiation comes from the U
region (which happens with probability $(\frac12
C_1+C_2)/(2C_F+C_A)$), only $p_2$ takes recoil, so that that
$p_2^{\out}\sim \Ko$, while $p_3^{\out}\ll p_2^{\out}$.
The part of the shift proportional to $C_2$ gets logarithmically
enhanced as in the previous case, while the one proportional to $C_3$
gives rise to a very singular contribution proportional to
$1/\sqrt{\alpha_s}$, coming from the average of $\ln Q/p_3^{\out}$ over
its DL Sudakov form factor $\exp\left\{-(\frac12
C_1+C_3)\frac{\alpha_s}{\pi}\ln^2Q/p_3^{\out}\right\}$. This gives
\begin{equation}
\label{eq:caso2}
\delta \Ko\sim 
 C_1\ln\frac{Q_1^{\mbox{NP}} e^{\eta_0}}{Q}+
C_2\ln\frac{Q_2^{\mbox{NP}}}{\Ko}
+C_3\frac{\pi}{2\sqrt{(\frac12 C_1+C_3)\alpha_s}}\>.
\end{equation}
The converse argument holds for an emission in the D region.
\end{itemize}
In conclusion, 
%\section{Conclusions}
we have now a promising theoretical method to deal with multi-jet
observables. Not only are we able to identify all sources of SL
contributions, but we can also relate the NP $1/Q$ power-suppressed
corrections with the ones which affect $e^+e^-$ two-jet shapes. A next
step will be the extension of the above results to event shapes
involving any number of jets. We hope that this analysis will greatly
improve the understanding of confinement physics, especially at hadron
colliders.

%%\paragraph{Acknowledgments}
\section*{Acknowledgements}
I am grateful to Andrea Banfi, Yuri Dokshitzer, Gavin Salam, Graham
Smye, Bryan Webber and Glulia Zanderighi which share the
responsibility for the results discussed on jet-shape analysis.

\end{document}